\documentclass[12pt]{article}
\usepackage{graphicx}
\usepackage{epsf,amsmath,bbold,amsfonts,stmaryrd}
\usepackage{mathrsfs}
\usepackage{appendix}
\usepackage{amssymb}
\usepackage{slashed}
\usepackage{float}
\usepackage{color} 
\usepackage[colorlinks]{hyperref}
\hypersetup{pageanchor=false,citecolor=red,urlcolor=red}
\usepackage{indentfirst}
\usepackage[numbers,square,comma,sort&compress,merge]{natbib}
\usepackage{multibib}
\usepackage[T1]{fontenc}
\usepackage[latin9,utf8]{inputenc}
\usepackage[english]{babel}

\def\D{\Delta}

\def\f{\frac}
\def\g{\gamma}

\def\G{\Gamma}

\def\l{\left}

\def\m{\mu}
\def\n{\nu}

\def\p{\partial}

\def\r{\right}
\def\s{\sigma}

\def\be{\begin{equation}}
\def\ee{\end{equation}}

\def\bea{\begin{eqnarray}}
\def\eea{\end{eqnarray}}

\def\ba{\begin{array}}
\def\ea{\end{array}}

\def\bc{\begin{center}}
\def\ec{\end{center}}

\def\bl{\begin{flushleft}}
\def\el{\end{flushleft}}

\def\br{\begin{flushright}}
\def\er{\end{flushright}}

\def\bi{\begin{itemize}}
\def\ei{\end{itemize}}

\def\bt{\begin{tabular}}
\def\et{\end{tabular}}

\newtheorem{question}{Question}
\def\bq{\begin{question}}
\def\eq{\end{question}}

\newtheorem{definition}{Def}
\def\bd{\begin{definition}}
\def\ed{\end{definition}}

\newtheorem{answer}{Answer}
\def\ban{\begin{answer}}
\def\ean{\end{answer}}

\newtheorem{possibleanswer}{Possible answer}
\def\bpa{\begin{possibleanswer}\normalfont}
\def\epa{\end{possibleanswer}}

\newtheorem{theorem}{Theorem}
\def\bth{\begin{theorem}}
\def\eth{\end{theorem}}

\begin{document}

\begin{titlepage}
\vspace{5cm}

\vspace{2cm}

\begin{center}
\bf \Large{Small--scale structure from neutron dark decay} 

\end{center}

\begin{center}
{\textsc {Georgios K. Karananas, Alexis
    Kassiteridis\,\footnote{Corresponding author.}}}
\end{center}

\begin{center}
{\it Arnold Sommerfeld Center\\
Ludwig-Maximilians-Universit\"at M\"unchen\\
 Theresienstra{\ss}e 37, 80333 M\"unchen, Germany}
\end{center}

\begin{center}
\texttt{\small georgios.karananas@physik.uni-muenchen.de} \\
\texttt{\small a.kassiteridis@physik.uni-muenchen.de} 
\end{center}

\vspace{2cm}

\begin{abstract}

It was recently proposed that the disagreement in the experimental
measurements of the lifetime of the neutron might be eradicated if the
neutron decays to particles responsible for the dark matter in the
Universe. In this paper we construct a prototype self--interacting
dark matter model which, apart from reproducing the correct relic
abundance, resolves all small--scale problems of the $\Lambda$CDM
paradigm. The theory is compatible with the present cosmological
observations and astrophysical bounds.

\end{abstract}

\end{titlepage}

\section{\label{sec:level1}Introduction}

Since the 30's, when Zwicky correctly claimed the existence of dark
matter (DM)~\cite{Zwicky:1933gu,Zwicky:1937zza}, a lot of things have
changed in the DM community. If one assumes that DM is made of
particles with a certain mass, then there are different approaches to
describe our universe: thermally produced Weakly--Interacting Massive
Particles (WIMPs), Feebly--Interacting Massive Particles (FIMPs) or
even warm DM (WDM), see for
instance~\cite{Asaka:2005an,Asaka:2005pn,Boyarsky:2009ix}. Towards the
end of the 20th century, the sole purpose of a DM candidate was to
solve the large--structure problems. However, critical questions have
arisen regarding the small--structure formation: the missing satellite
problem,\footnote{WDM models are usually able to solve the abundance
problem of dwarf galaxies by employing particles with masses in the
keV range~\cite{Bringmann:2016ilk}. On the other hand, cold DM (CDM)
theories require late kinetic decouplings $T\sim~\text{keV}$, in order
to provide a possible
solution~\cite{Vogelsberger:2015gpr,Bringmann:2016ilk,Tulin:2017ara}. }
the cupsy profiles of dark galaxies in the $\Lambda$CDM cosmology and
the fact that these galaxies are too big to fail in producing luminous
content; for a thorough review and possible solutions of these
problems
see~\cite{Tulin:2017ara,Vogelsberger:2015gpr,Vogelsberger:2015gpr,
Schewtschenko:2015rno}.  The DM model--building ``industry'' is
usually dedicated to the pursuit of theories which produce the right
DM relic density, while solving all small--scale problems in the local
group.

The idea that DM can be naturally accomodated within a ``hidden
world'' that communicates with the Standard Model (SM) of particle
physics via a neutron portal, is certainly not a new one. Actually,
this possibility was first spelled out, in the context of braneworld
scenarios, by Dvali and Gabadadze~\cite{Dvali:1999gf}. Nevertheless,
owing to the fact that the aforementioned portal is omnipresent in
theories with dark sectors (DS)~\cite{Dvali:1999gf}, there is a number
of interesting implications. For instance, in~\cite{Dvali:2009fw}, it
was shown that if the hidden sector involves a big number of SM
copies, then the DM abundance can be naturally produced, since the
particle content of the other copies are excellent DM
candidates~\cite{Dvali:2009ne}. Recently, the idea that DM might be
due to the interaction of the neutron with a DS has been revived and
studied in great detail in~\cite{Fornal:2018eol}; various aspects of
this proposal have been addressed in a number of follow--up
works~\cite{Pitrou:2018cgg, Tang:2018eln,Czarnecki:2018okw,
Serebrov:2018mva,McKeen:2018xwc, Baym:2018ljz,
Motta:2018rxp,Cline:2018ami,Gonzalez-Alonso:2018omy,Sun:2018yaw,
Leontaris:2018blt,Ejiri:2018dun}.

A new dark decay channel for the neutron can resolve the experimental
discrepancy---which is more than $4\s$---between the bottle and beam
measurements of the neutron lifetime.  The neutron lifetime inferred
from the $\beta$--decays of free neutrons in a beam is found to be
$\tau_n^{\beta^-} \approx 888~\text{s}$~\cite{Yue:2013qrc}. At the
same time, the direct counting of the remaining neutrons in the bottle
experiment leads to $\tau_n \approx
880~\text{s}$~\cite{Patrignani:2016xqp}. If yet another channel for
neutron decay is present, then the correct lifetime is the one
measured in the bottle experiment, meaning that its decay rate is
\begin{equation}
\Gamma \equiv \frac{1}{\tau_n} = \Gamma_{\beta^-} + \Gamma_{\rm{DS}}\ ,
\end{equation}
 where $\Gamma_{\beta^-} = 1/ \tau_n^{\beta^-} $, while
 $\Gamma_{\rm{DS}}$ corresponds to the decay channel into the DS
 particles. It is straightforward to show that~\cite{Fornal:2018eol}
\begin{equation} 
\label{neut_dec_dark}
\Gamma_{\rm{DS}} \approx 10^{-5} \, \rm{s}^{-1}\  .
\end{equation}

Kinematically, the neutron decay to the DS is allowed if the sum of
the masses of its decay products, $\sum m$, is smaller than the mass
of the neutron, $m_n=939.565~\text{MeV}$. On the other hand, in order
to be compatible with all present experimental constraints due to the
stability of the proton, as well as nuclear transitions between stable
nuclei, $\sum m$ must be bigger than
$937.900~\text{MeV}$~\cite{McKeen:2015cuz, Fornal:2018eol}. Let us
note that, in principle, the hidden modes could decay back into (SM)
protons, therefore, we consider only
\begin{equation} \sum m < m_p+m_e+\min[m_\nu]\ ,
 \end{equation} where $\min[m_\nu]$ denotes the smallest of the
neutrino masses. In other words, only a very small window for the
allowed mass--sum of the DM particles is open
\be
\label{mass interval}
937.900~\text{MeV} <\sum m < 938.783~\text{MeV} \ .
\ee

The main aim of this paper is to show that the products of the
possible hidden decays of the neutron---that could successfully
explain its lifetime anomaly---not only give rise to the DM relic
abundance, but might be able to simultaneously provide solutions to
the enduring small--scale structure problems of the $\Lambda$CDM
cosmology. As a proof of concept, we construct a self--interacting
dark matter (SIDM) model~\cite{Spergel:1999mh}, which is compatible
with all astrophysical and particle physics constraints and respects
all known symmetries of the SM.  To the best of our knowledge, up to
now there has not appeared a theory in which the products of the
hidden neutron decay are able to alleviate all the small--scale
problems.

In our setup, the present DM abundance is assumed to consist of a
Dirac fermion with mass in the GeV range---charged under a hidden
abelian gauge symmetry---that plays the role of the stable DM
candidate.  Both the SIDM virtue of the model, as well as its late
kinetic decoupling, is due to the secluded interaction mediated by a
vector. Note that this is the only choice of fields that allows the
theory to accomodate repulsive self--interactions, see for
example~\cite{Tulin:2013teo,Tulin:2017ara,Cline:2018ami}. It should be
stressed that, contrary to what stated in~\cite{Cyr-Racine:2015ihg},
Yukawa self-interactions mediated by a scalar are \emph{always
attractive}, irrespective of whether the involved fermions are Dirac
or Majorana.\footnote{\label{foot:attract}We thank Jonathan Cornell
for bringing this to our attention, which in turn led to the revision
of our paper.}

To ensure the compatibility of the model with the various constraints,
it is nevertheless necessary to introduce more degrees of
freedom. These comprise a heavy Dirac fermion whose mass needs to be
of the order of TeV, that opens the hidden decay channel of the
neutron and at the same time modifies the WIMP production of the DM
particles, as well as a very light complex scalar field. Both of them
should admit a baryon number, while the scalar is also charged under
the hidden symmetry.

This paper is organized as follows. In Sec.~\ref{sec:eff_theory}, we
introduce the model and discuss its field content. In
Sec.~\ref{sec:constraints}, we compute certain important observables
(decay rates and cross sections) and we give a complete list of
constraints stemming from particle physics and cosmological
considerations. In Sec.~\ref{sec:therm_ev}, we study the thermal
evolution of the theory by computing the present--day DM abundance and
the kinetic decoupling temperature. In Sec.~\ref{sec:small-scale}, we
examine how the small--scale problems are solved successfully in this
context. Our conclusions can be found in Sec.~\ref{sec:concl}.

\section{The effective theory for neutron dark decays} 
\label{sec:eff_theory}

The proposed model is an effective theory which is valid below some
cutoff scale $\Lambda$ and connects the neutron hidden decay with an
invisible sector that plays the role of DM. The most general
color--singlet operators that open up such a channel read
\be
\label{neut_op_decay}
\f{1}{\Lambda^{2+N}}\, \bar\Psi_N P_s d\,  u^\text{T} C P_s d
+\text{h.c.} \ , 
\ee
with $\Psi_N$ some fermionic operator (fundamental or composite) that
carries unit baryon number.  Here, $P_s~\text{and}~C$ denote the
chiral projection and charge conjugation operators in the
spinor--space, respectively, and $s$ is running over the
chiralities. To keep the notation simple, we have suppressed the
$SU(3)$--color labels.

The thermal evolution of the theory takes place at temperatures $T\ll
\Lambda$, thus the communication of the SM with the dark sector is due
to the lowest--order effective neutron portal (corresponding to $N=0$),
i.e.
\be
\label{neutronphil portal}
\frac{1}{\Lambda^2}\, \bar{\Psi}_0 P_s d \, u^{\rm T}C P_s d +
\text{h.c.}\ .
\ee
Here, $\Psi_0$ is a Dirac fermion which is a singlet under the SM
gauge group, but charged under the (global) baryon symmetry; in what
follows we will call it heavy baryon.  For our considerations, higher
order terms can be safely neglected at these energies.\footnote{Note
that a possible ultraviolet completion of the effective
interaction~(\ref{neutronphil portal}) can certainly be found. For
instance, one can introduce a colored scalar field in an appropriate
representation of the SM gauge group (see
e.g.~\cite{McKeen:2015cuz,Fornal:2018eol}), and mass of the order of
$\Lambda$.}

As it will become clear in a while, in order to have a
phenomenologically viable DM model, the spectrum of the hidden sector,
in addition to $\Psi_0$, should comprise: \emph{i)} a stable Dirac
fermion $F$ with mass $m_F\lesssim\mathcal O(\text{GeV})$, which is a
singlet under the SM gauge group and does not carry baryon number, it
is, however, charged under a hidden local abelian symmetry with charge
$g$; \emph{ii)} a complex scalar field $\Sigma$ with mass
$m_\Sigma\sim\mathcal O(\text{eV})$, unit baryon number and secluded
charge $-g$, which we call scalar baryon; \emph{iii)} a real vector
field $\s$ with $m_\s\sim \mathcal O(\text{keV})$ and no baryon
number, acting as the force mediator of the hidden gauge
symmetry. 

In this work we are agnostic as to how the participating fields
acquire their masses. It should be stated clearly that we are
interested in working out a purely phenomenological model for the dark
decay of the neutron, which successfully addresses the DM
problems. There is no a priori reason for the mass hierarchy to be
chosen like this, besides fulfilling the various observational
constraints. For instance, the scalar baryon $\Sigma$ could, in
principle, communicate directly with the SM sector through a quartic
interaction with the Higgs field. However, the corresponding coupling
should be finetuned to be extremely small, even for a feebly
production DM mechanism; otherwise, the desired value for $m_\Sigma$
will not be generated. In any case, such interactions are irrelevant
regarding the thermal evolution of the proposed model, so we will not
discuss them further in the present paper.

The \emph{cosmologically relevant} interaction terms appearing in the
hidden sector are the following (see also eq. (B26)
in~\cite{Bringmann:2016ilk})
\begin{equation}
\label{dsinte}
\mathscr{L}_\text{hidden} \supset - \lambda\left(\Sigma\,
\bar{\Psi}_0 F + \text{h.c.} \right) +g \, \s^\m\, \bar{F}\g_\m F -
\mathrm{i} {g}\, \sigma^\mu \,\Sigma^*\overleftrightarrow{\p
_\m}\Sigma \ ,
\end{equation}
with $\lambda~\text{and}~g$ dimensionless couplings and 
$\Sigma^*\overleftrightarrow{\p _\m}\Sigma\equiv\Sigma^* \p_\m \Sigma
-\Sigma \p_\m \Sigma^*$.

Let us now turn to the justification of the terms appearing
in~(\ref{dsinte}). The first one, apart from opening up the desired
dark decay channel for the neutron (if of course the mass spectrum
allows it), it is also responsible for the freeze--out annihilation of
the DM particles.  The second, enables a SIDM scenario and gives rise
to possible infrared--dominant interactions in the DM sector of the
theory. Such interactions are vital for an acceptable small--scale
structure formation~\cite{Tulin:2017ara}. Finally, the third term
in~(\ref{dsinte}), makes $\s$ potentially unstable, since they could
decay mainly in scalar baryons. In other words, the scalar baryons
serve as way--out particles and a possible overclosure of the universe
at late times is avoided.\footnote{In principle, one could also allow
an additional effective coupling between $\sigma$ and the
SM--neutrinos $\n$ of the form $\s^\m\, \bar{\n}\g_\m \g_5\n$\,; such
couplings are discussed in~\cite{Balducci:2017vwg}.} For later
convenience, the decay width is given by
\begin{equation} 
\label{gam_sigm_2Sigm}
\Gamma_{\sigma \rightarrow 2\Sigma} \approx
\frac{\alpha'}{12 }\,  m_\sigma
\end{equation} 
 with $\alpha'= g^2/4\pi$, and assuming $m_\sigma \gg m_\Sigma$. In principle the scalar baryon can directly couple to the SM-Higgs field through a gauge invariant quadratic-scalar interaction with some coupling $\lambda_\Sigma$. However, $m_\Sigma$ should lie at the eV or sub-eV in order to avoid a possible overclosure; this leads to extremely small values for the  corresponding coupling $\lambda_\Sigma$, making such interaction terms cosmologically irrelevant (even for a FIMP scenario).

The model under consideration admits all the properties of a prototype
SIDM theory, due to the presence of the mediator. The SIDM cross
sections per dark matter mass are repulsive~\cite{Tulin:2013teo,
Tulin:2017ara} and are strongly velocity--dependent when $2 \alpha'  
m_\s/ m_Fv_{\rm rel}^{\; 2} \lesssim
10^3$~\cite{Tulin:2017ara}. This fact is in principle essential for
addressing the small--scale problems that are present when matter
structures form in non--interacting (purely WIMP) models. Furthermore,
it is this very SIDM interaction that enables the observed neutron
star formation and makes the proposed theory compatible with the
latest observations.

\section{Constraints from particle physics and cosmology}
\label{sec:constraints}

In this section we discuss the possible constraints on the couplings
and masses of the effective theory. For an extensive list of
constraints on interactions between SM--particles and DM,
see~\cite{Shoemaker:2013tda}.

\subsection{Particle and astroparticle physics}

The motivation behind this model is purely phenomenological and aims
in connecting the existence of a hidden decay channel for the neutron
to a dark sector which addresses all the DM problems. As we already
mentioned, for the dark neutron decay channel to be available, we have
to require that $\sum m \equiv m_F+m_\Sigma$ should lie in the allowed
mass interval~\eqref{mass interval}, and that
\begin{equation}
m_{\Psi_0}\gg  m_F+m_\Sigma \ .
\end{equation}
The validity of the effective field theory approach dictates that
\begin{equation}
 m_{\Psi_0} < \Lambda\  .
 \end{equation} 
At temperatures below the quark confinement scale $\Lambda_{\rm QCD}
\approx 200$ MeV, the neutron portal~\eqref{neutronphil portal} boils
down to
\begin{equation} 
\label{neutron portal}
\frac{f_n}{\Lambda^2}\bar{\Psi}_0 n + \text{h.c.}\, ,
\end{equation}
where $f_n \approx 10^{-2}$ GeV$^3$ is the neutron decay
constant~\cite{Aoki:2017puj}.  After diagonalizing the mass matrix and
upon integrating out the heavy $\Psi_0$, the hidden decay channel
arises from the following effective Yukawa term
\begin{equation}
y\, \Sigma\, \bar{n}  F +\text{h.c.} \ ,
\end{equation}
where the dimensionless coupling is defined as $y \equiv \lambda
f_n/\Lambda^2m_{\Psi_0}$, and in abuse of language $n$ and $F$
correspond to the mass eigenstates with eigenvalues $m_n$ and $m_F$,
respectively.

From the above interaction term, the invisible two--body decay width
of the neutron can be easily calculated. At the lowest order in the
coupling constant $y$, one finds
\be
\G_{\rm{DS}}\approx \f{y^2}{8\pi
m_n^3}\l[(m_n+m_F)^2-m_\Sigma^2\r]\l[(m_n^2-m_F^2-m_\Sigma^2)^2
-4m_F^2m_\Sigma^2\r]^{1/2} \ .
\ee
In order to reproduce the desired value~(\ref{neut_dec_dark}), the
coupling constant should be $y\sim \mathcal O(10^{-13})$. This means
that $y$ is actually completely fixed via the neutron anomaly and is
not a free parameter. However, if $y$ were to be complex, then this
effective interaction could in principle modify the electric dipole
moment of the neutron. Nevertheless, due to the tiny value of $y$, the
corresponding correction should not lead to additional constraints.

The DM particles $F$ are assumed to be stable and produced in local
thermal equilibrium (LTE). Furthermore, possible upper bounds on the
dark matter mass~\cite{Griest:1989wd} are irrelevant due to
$m_F\sim\mathcal{O}(1)$~GeV. Lower bounds on the boson mass are not
valid since the dark radiation (or in other words the $\s$ and
$\Sigma$ particles) is not in LTE with the SM modes during the typical
neutrino decoupling at $T_{\nu D} = 2.3$
MeV~\cite{Enqvist:1991gx}. The low--temperature annihilation of $F$
into scalar baryons implies that $m_F>m_\Sigma$.

The presence of the effective interaction~(\ref{neutron portal}),
between $F$ and the neutrons leads to a 2--to--2 interaction with
spin--independent cross section
\begin{equation}
\sigma_{nF\to nF}\sim \f{\Gamma_{\rm{DS}}^2}{(vm_n)^4} \sim
\mathcal{O}(10^{-69})\rm{cm}^2\ ,
\end{equation}
where the characteristic velocity of the scattering is $v\sim \mathcal
O\l(10^{-3}\r)$ in units of speed of
light~\cite{Akerib:2016vxi}. Similar considerations apply also to the
spin--dependent cross sections. These are totally in line with the LUX
and XENON direct detection measurements for WIMP/neutron elastic
scattering~\cite{Hamze:2014wca,Akerib:2016vxi,Chao:2016lqd,
Aprile:2017iyp}.

Finally, since the dark sector does not break the baryon symmetry,
constraints inferred from baryon-violating processes such as the
($\Delta B=2$) $n-\bar{n}$ oscillations \cite{Abe:2011ky}, and
$2n\rightarrow2\pi$ dinucleon decays into pions
\cite{Gustafson:2015qyo}, are not applicable here. If, on the other
hand, $\Sigma$ were real, then the relevant parameter space of the
proposed theory would be excluded. It should be stressed at this point
that the $\Psi_0$ is uncharged under the hidden local symmetry, which
is of vital importance in order to obtain a less constrained parameter
space.

\subsection{Cosmology}

In order to solve the small--scale issues of the $\Lambda$CDM
paradigm, $m_\s$ should lie in the sub--MeV region. Such values give
rise to interesting effects at the dwarf and cluster galaxy scales and
could allow a late kinetic decoupling of $F$ from the plasma. However,
much lower values of $m_\s$, i.e. sub--keV masses tend to oversolve
the small--scale problems of $\Lambda$CDM cosmology leading to
unacceptably large protohalo masses, while the corresponding large
SIDM cross sections destroy any core structures at dwarf and cluster
scales~\cite{Vogelsberger:2015gpr}.

Since $m_\s<\mathcal O(\text{MeV})$, these modes are in LTE and
ultra-relativistic during the Big Bang Nucleosynthesis (BBN). In
principle they could modify the processes that take place at this
period, and consequently affect the energy density of the universe
after the neutrino decoupling. This is encapsulated in the deviation
of the effective neutrino degrees of freedom $\D N_{\rm eff}$.  For
the proposed theory, we find that the dark sector decouples from the
SM plasma before the annihilation of the bottom quark. This leads to
an entropy dilution $\left(T_{\rm DS}/T_{\n}\right)^3\vert_{T_{\n
D}}\sim 0.2$, in line with BBN~\cite{Cyburt:2015mya} and
CMB~\cite{Ade:2015xua} 1$\sigma$--measurements. Such a value
corresponds to $\Delta N_{\rm eff} \vert_{\rm BBN} \approx 0.35$,
while during recombination $\Delta N_{\rm eff} \vert_{\rm CMB} \approx
0.15$, for $m_\sigma \sim~\text{keV} $ and $m_\Sigma\sim~\text{eV}$,
which are perfectly compatible with all present measurements. At the
same time, the scalar baryons comprise less than $2\%$ of the total
CDM abundance.  Moreover, this result may also explain the recent
tension about the difference between $\Delta N_{\mathrm{eff}}$ at BBN
and CMB periods. In our model, $\Delta N_{\rm eff}\vert_{\rm CMB}-
\Delta N_{\rm eff}\vert_{\rm BBN}<0$, appears naturally, since the
force mediators become nonrelativistic after the BBN period but long
before the era of recombination. Note that keeping $\Sigma$ massless
or $m_\Sigma \lesssim~0.3~\text{eV}$, leads to an increase of $\Delta
N_{\rm{eff}}=0.49$, which nevertheless is still 2$\sigma$--compatible.

In order to obtain the thermal evolution of the theory, where all
participating modes were in LTE at previous times and still be
compatible with the effective description we proposed, we should
demand that the heavy baryons be in LTE with the light quarks at
temperatures below the cutoff $\Lambda$. It turns out that this is
always the case as long as $\Lambda\ll M_{\rm Pl}= 1.22\times
10^{19}~$GeV and $y \sim \mathcal{O}(10^{-13})$, which directly
relates a hidden neutron decay to a non--zero relic abundance.
Meanwhile, the decoupling of the DS from the SM plasma should take
place well before the QCD phase transition after which the
relativistic degrees of freedom reduce drastically.  Otherwise, the
extracted value of the entropy ratio would lead to unacceptably large
deviations of effective neutrino degrees of freedom, excluded from
Planck satellite measurements. The above reasoning yields $1\,
{\rm{TeV}} \lesssim \Lambda \ll M_{\rm Pl}$.
 
Recent constraints~\cite{McKeen:2018xwc,Baym:2018ljz,Motta:2018rxp,
Cline:2018ami}, about the hidden sector communicating with the
neutron, are inferred from the observed masses of neutron stars
assuming WIMP theories. If the repulsive (vector--mediated)
self--interactions between $F$ are strong inside the star, then the
equation of state is modified and the bounds might be
evaded~\cite{McKeen:2018xwc}. Since the theory under consideration has
all the virtues of a SIDM scenario, the observations of neutron stars
with size of the order of $2M_{\odot}$, can in principle be
accommodated. More precisely, the coupling constant $\alpha'$ should
lie in the milli--regime, while the mediators should admit keV masses,
yielding~\cite{McKeen:2018xwc,Cline:2018ami}
\begin{equation}
\label{eq:charg_bound_NS}
\alpha' \gtrsim
\mathcal{O}(10^{-11})\left(\frac{m_\s}{\text{keV}}\right)^2\ . 
\end{equation}

In addition, for such values of $m_\sigma\sim \mathcal{O}(\text{keV})$
and $\alpha'\sim \mathcal{O}(10^{-4})$, and if the DM mass lies well
below the GeV regime, then in principle dark halos could appear around
neutron stars, as shown and explained in~\cite{Nelson:2018xtr}. Such
non-compact objects imply further constrains on the repulsive DM
self-interactions. As explained previously, the DM mass is almost
fixed and lies in the small interval~(\ref{mass interval}), namely
$m_F \sim \mathcal{O}(\text{GeV})$; this relaxes the proposed dark
halo constraints on $\alpha'$ for this particular $m_\sigma$.

\section{Thermal evolution of the theory}
\label{sec:therm_ev}

In this part of our work we study the thermal history of the proposed
theory. We calculate the dark matter relic abundance, which consists
of the $F$--modes after they chemically decouple from the plasma and
we examine the possibility of a late kinetic decoupling regime. All
temperatures that appear here are given in the reference frame of the
photons, unless stated otherwise.

\subsection{The present--day relic density}

The fact that $m_F\sim \mathcal O(\text{GeV})$ and $m_\sigma \sim
\mathcal{O}(\rm{keV})$, implies that the SIDM couplings are much
smaller than the usual WIMP couplings. Consequently, it is safe to
assume that $\lambda \gg g$. In turn, for temperatures $T<m_F$, this
leads into the rapid dominant annihilation of $F$ into the scalar
baryons. This process is mediated through the heavy baryon $\Psi_0$,
as long as $m_\Sigma \ll m_F$. The corresponding thermal cross section
reads
\begin{equation} 
\langle v_{\rm rel} \sigma_{\rm ann} \rangle_{F\rightarrow \Sigma}
=\frac{\pi\alpha^2}{4 m^2_{\Psi_0}} \langle v_{\rm rel}^2 \rangle \ , 
\end{equation}
where $\langle\ldots\rangle$ denotes the thermal average using the
relative velocities $v_{\rm rel}$, and for later convenience we
introduced $\alpha\equiv \lambda^2/4\pi$. Assuming that the $F$'s
dominate the DM population, their annihilation into $\Sigma$ reduces
their number. This fixes the coupling constant $\alpha$ for a given
set of masses reproducing the measured DM relic density $\Omega_{\rm
CDM}$.

In what follows, we undertake the approach outlined
in~\cite{Balducci:2017vwg} and~\cite{Alex}.  The Boltzmann equation
for the distribution function $f_F(t,\textbf{p})$ of the DM particles
$F$ in a Friedmann--Lema{\^i}tre--Robertson--Walker universe reads
\be
\label{boltz_equation}
\f{\p f_F}{\p t}-3H \textbf{p} \cdot \f{\p f_F}{\p \textbf{p}}=
\frac{1}{2p^0}\, C \ 
, 
\ee
where $H$ is the Hubble parameter and $C$ the collision integral, whose
formal expression can be found for instance in~\cite{Kolb:1990vq} and
\cite{Bringmann:2016ilk}. Note that the collision term can be
simplified by assuming that the final states follow an equilibrium
thermal distribution. As customary, it is quite convenient to
eliminate the Hubble parameter from~(\ref{boltz_equation}), by working
in terms of the number density per comoving volume, and study its
evolution with respect to $m_F/T.$

The resulting differential equation is solved numerically by setting
the initial conditions at the chemical
freeze--out~\cite{Hofmann:2001bi,Balducci:2017vwg,Alex}. This yields
the present--day relic abundance of $F$ as a function of $m_{\Psi_0}$
and the coupling constant $\alpha$
\begin{eqnarray}
	\label{om}
\Omega_{{F}} h^2 \approx \f{0.12}{2}\;
\left(\frac{\alpha}{0.1}\right)^{-2}\; \left(\frac{m_{\Psi_0}}{0.8\,
{\rm TeV}} \right)^2\ ,
\end{eqnarray}
with $h\approx 0.67$, the reduced Hubble constant~\cite{Ade:2015xua}.
One notices immediately the absence of $m_F$ in the DM relic density
at the first order approximation. This is an aftermath of the presence
of the much heavier field $\Psi_0$ in the spectrum of the theory,
which acts as the mediator during the annihilation process.

\subsection{Kinetic decoupling at the keV--scale}

Up to this point, we have shown that the theory is capable of
explaining the anomaly of the neutron lifetime, and at the same time
producing the desired relic abundance of DM. However, the solution to
the small--scale ``crisis'' lies also in the elastic scattering
between $F$ and the scalar baryons in the dark sector.  The kinetic
decoupling temperature, henceforth $T_{\rm kd}$, describes the moment
when these processes cease to sustain LTE~\cite{Bringmann:2006mu}.

The phenomenologically interesting case is when the kinetic decoupling
takes place long after the BBN period $(T_{\rm
BBN}\sim\text{few~MeV})$, in order to suppress the structure formation
at scales similar to those of dwarf galaxies.  For a novel approach of
the thermal kinetic decoupling regime see~\cite{Bringmann:2006mu}, and
for a detailed analysis of theories that enable such a late decoupling
see~\cite{Bringmann:2016ilk}.

We assume that $F$ at high temperatures are in LTE with the scalar
baryons. As the temperature falls below $m_\sigma$, these modes
annihilate and decay into scalar baryons. To estimate $T_{\rm kd}$, we
have to equate the Hubble parameter $H(T)$ to the elastic scattering
rate $\G_{\rm el}(T)$, which is given by
\begin{equation}
\label{gamma_el}
\Gamma_{{\rm el}}(T)\approx \frac{2}{3\pi^2 m_F}\int^\infty_0
\mathrm{d}E\, f_{\Sigma}(E) \frac{\partial}{\partial
{E}}\left(E^4\sigma_{ \rm{el}}\right) \ .
\end{equation}
As usual, $ f_{\Sigma}(E)$ is the equilibrium distribution function
per d.o.f. of the scalar baryons at temperature $T$, $E$ its energy,
and $\s_{\rm{el}}$ the momentum--transfer elastic cross section, given
in~\cite{Bringmann:2016ilk}.

At temperatures $T\ll m_F $, and at the lowest order in perturbation
theory, one obtains for the IR--dominant expression
\begin{equation}
\sigma_{\rm{el}}
\approx\frac{4\pi \alpha'^2}{ E^2}\left[\log
\left(\frac{4E^2+m_\s^2} {m_\s^2}\right)
-\frac{E^2}{E^2+\frac{1}{4}m_\s^2}\right]\, .
\end{equation}
Note that as the temperature falls below $m_\s$, the elastic
scattering becomes dominant. Here, we ignored the optical term due to
the decay of the vector field since $\Gamma_{\sigma\rightarrow
2\Sigma}\ll m_\sigma$, see~(\ref{gam_sigm_2Sigm}).

Before concluding this section, let us compute the kinetic decoupling
temperature for certain values of the relevant parameters. For
example, considering the benchmark point $m_\sigma \approx 10$ keV and
milli--charges of $\alpha' \approx 3 \times 10^{-8}$
(c.f. equation~(\ref{eq:charg_bound_NS})) we find
\be
\label{eq:temp_kd}
 T_{\rm kd} \approx 0.45~\text{keV} \ . 
\ee
In principle, much smaller values of $m_\sigma$ or/and larger charges,
give rise to lower decoupling temperatures leading to unacceptable
results as explained previously.

\section{The small--scale structures}
\label{sec:small-scale}

The aim of this section is to show that the products of possible
hidden decays of the neutron can solve the DM small--scale structure
problems. Therefore, we discuss the implications on the small--scale
structure formation due to the late kinetic decoupling, together with
the SIDM nature of the model. Our analysis is based
on~\cite{Balducci:2017vwg} and~\cite{Alex}.

\subsection{The protohalo mass: the abundance of satellite galaxies}    

The kinetic decoupling is of great importance for the small--scale
structures: the efficient momentum exchange damps the perturbations
and determines the masses of the first gravitationally--bound objects
(protohalos) of DM
particles~\cite{Green:2005fa,Shoemaker:2013tda}. This process
terminates after $T_{\rm kd}$. Technically, the fluctuations are
exponentially damped with a characteristic mass scale given by
\be
\label{mass_scale_damp}
M_{\rm damp}=\f{4\pi}{3} \f{\rho_{\rm m}(T_{\rm kd})}{H^3(T_{\rm kd})} 
\ ,
\ee
with $\rho_{\rm m}(T_{\rm kd})$ the matter density at the time of the
kinetic decoupling.  In other words, the dark matter substructures
cannot form within the Hubble volume at kinetic decoupling, since the
scalar baryon abundance within this volume suffices to keep $F$ in
approximate LTE. As an example, masses of order $M_{\rm
damp}\sim 10^8M_{\odot}$ correspond to $T_{\rm kd}$ of the order of
keV.\footnote{We do not consider the free--streaming impact due to the
much heavier DM particles appearing in this theory as compared to the
usual WDM mass interval.}

The small-scale problems of the $\Lambda$CDM cosmology can be
addressed more efficiently after suppressing the linear power spectrum
at scales similar to those of dwarf
galaxies~\cite{Vogelsberger:2015gpr,Aarssen:2012fx,Schewtschenko:2015rno}.
This means that damping masses between $10^8-10^9$ solar masses are
required. Such cutoffs are generated when the positrons kinetically
decouple from their scattering partners at $T_{\rm kd}
\sim\text{keV}$, as stated in~\cite{Tulin:2017ara,Lovell:2017eec}. On
the other hand, the cutoff masses should not be larger than the bounds
set by Lyman--$\alpha$ measurements
~\cite{Aarssen:2012fx,Shoemaker:2013tda,Bullock:2010uy,Baur:2015jsy,
Irsic:2017ixq}.

The proposed theory provides naturally $T_{\rm kd} \sim\mathcal O(
m_\sigma)$, and therefore cutoff masses of the desired order are
easily accessible, while respecting all present constraints.  For
comparison, we note that CDM theories, where no late kinetic
decoupling is present, usually predict masses between the Earth
mass~\cite{Green:2005fa} and below the mass of Sagittarius
A*~\cite{Bringmann:2016ilk}.

\subsection{Self--interaction scattering in the dwarf and cluster
  scales} 

In this last part of the present work, we study the impact to the
small--scale structure formation at the non--linear regime due to the
protohalo masses and the SIDM properties of the theory. In other
words, we examine whether this model is able to solve successfully the
enduring small--scale problems of the $\Lambda$CDM paradigm: the cusp
vs. core and the too big to fail problems. Moreover, if one considers
also the missing satellite issue as an actual problem, then it is
already solved due to the produced damping masses discussed
previously. However, we tend to understand the solution of this
problem more as a virtue of the underlying theory helping to address
the cusp vs. core and the too big to fail problems more easily.

It is well known that a natural solution to both of these problems can
be found in the context of SIDM~\cite{Tulin:2017ara}. For this to be
possible, the elastic cross-section per unit mass should at least be
$\langle\sigma_T/m\rangle_{v_{\rm therm}}\sim 1$cm$^2$g$^{-1}$, where
the thermal average is taken with respect to a Maxwell-Boltzmann
distribution, with $v_{\rm therm}$ as the most probable velocity.
However, due to the typical velocities of the DM particles on cluster
scales, it turns out that $\langle\sigma_T/m\rangle_{v_{\rm
therm}}\sim
0.1$cm$^2$g$^{-1}$~\cite{Tulin:2017ara,Bondarenko:2017rfu}, which
implies that the SIDM elastic cross section should have a (mild)
velocity dependence.\footnote{This is also the case for
neutron--proton scattering in the SM.} If we ignore the damping of the
power spectrum due to the kinetic decoupling, then the above values
seem to be able to resolve the cusp vs. core and the too big to fail
problems as shown in~\cite{Vogelsberger:2012ku,Zavala:2012us} and
explained in~\cite{Tulin:2017ara}. On the other hand, due to the
combination of the SIDM effects together with the late kinetic
decoupling, one has to make sure that the aforementioned problems are
not actually over-solved.

For the model that we discussed here, it is easy to obtain
$\langle\sigma_T/m\rangle_{v_{\rm therm}}\sim (0.1-10) \, {\rm cm}^2
{\rm g}^{-1}$ ~\cite{Bringmann:2016din}, as long as $m_\s \sim
\mathcal O(\text{keV})$ and $\alpha' \sim 10^{-8}$. For instance, we present
the following benchmark point: for thermally produced $F$ with
$m_F=937.9$ MeV, $m_{\Psi_0} = 0.8$ TeV, $\Lambda \approx 14$ TeV and
$m_\s=10$ keV, a late kinetic decoupling takes place at $T_{\rm kd}
\approx 0.45~\text{keV}$ (see eq.~(\ref{eq:temp_kd})). At the same
time, the thermally--averaged SIDM elastic cross sections are similar
to the ones of the tuned framework
ETHOS--4~\cite{Vogelsberger:2015gpr}. This model is perfectly
compatible with the latest cosmological data, alleviates the missing
satellite issue, and solves the too big to fail and the cusp vs. core
problems as well~\cite{Lovell:2017eec}, without over--solving them.

\section{Conclusions}
\label{sec:concl}

It was recently suggested that the large discrepancy in the measured
neutron lifetime from the different experiments may be due to the
presence of a hidden sector, which sources the present--day DM relic
density. However, apart from reproducing the appropriate $\Omega_{\rm
DM}$, a theory describing DM should also address the enduring
small--scale problems of the $\Lambda$CDM cosmology, while staying in
line with all recent experiments.

The main aim of this paper was to provide an existence proof of a
theory in which the possible dark decay products of the neutron solve
the DM problems in the local group. More precisely, they admit the
observed DM relic density and at the same time address the enduring
small--scale structure problems of the $\Lambda$CDM paradigm, while
staying compatible with all astrophysical and particle physics
constraints.

In our context, a late decoupling takes place naturally for mediator
masses at the keV scale. Furthermore, it is this very mediator that
enables magnitudes of the corresponding SIDM cross sections. These are
not only necessary in order to solve the enduring small--scale
problems of the concordance cosmology, but also to generate
sufficiently heavy neutron stars compatible with all observations,
while reducing the recent tension about the effective neutrino degrees
of freedom. The fact that the parameter space favors masses of order
$\mathcal{O}(1-10)$ keV, might be directly related to experimental
evidence: a few years ago a mysterious 3.55 keV photon ray was
measured~\cite{Bulbul:2014sua,Boyarsky:2014jta}, indicating the
possibility of the existence of lighter modes in the Universe.

\section*{Acknowledgments}

We thank Ottavia Balducci, Javier Rubio and especially Gia Dvali and
Stefan Hofmann for discussions and suggestions. We are grateful to
Jonathan Cornell for pointing out to us an incorect statement in a
previous version of the manuscript concerning the nature of
scalar--mediated self--interactions for Majorana fermions
(footnote~\ref{foot:attract} and related discussion in the
Introduction). G.K.K. is grateful to the Theoretical High Energy
Physics Group of the National Technical University of Athens, where
part of this work was carried out, for warm hospitality. This work was
supported by the ERC-AdG-2013 grant 339169 ``Selfcompletion,'' the DFG
cluster of excellence ``Origin and Structure of the Universe,'' the
Humboldt Foundation and the TRR 33 ``The Dark Universe.''

\bibliographystyle{utphys}
\bibliography{neutron_DM}{}

\end{document}